# Relationship between electronic and crystal structure in $Nd_{1+x}Ba_{2-x}Cu_3O_{6+\delta}$

A.V. Fetisov [a,*], S.Kh. Estemirova [a,b]

[a] *Institute of Metallurgy of the Ural Branch of the Russian Academy of Sciences,
101 Amundsen Str., 620016 Ekaterinburg, Russia*

[b] *Ural Federal University, 19 Mira St., 620002 Ekaterinburg, Russia*

**Abstract**

In accordance with a relation between the lattice parameter $c$ and the concentration of holes $p$, which has been found earlier for the $YBa_2Cu_3O_{6+\delta}$ compound, we obtain the $p$ parameter in the same manner for the solid solution $Nd_{1.2}Ba_{1.8}Cu_3O_{6+\delta}$ with substitution of $Nd^{3+}$ for $Ba^{2+}$ in barium sites. For this, structural and superconducting properties of the $Nd_{1.2}Ba_{1.8}Cu_3O_{6+\delta}$ samples stored for specified times $\tau$ after quenching from annealing temperatures have been studied. The investigation shows that $Nd_{1.2}Ba_{1.8}Cu_3O_{6+\delta}$ stored for a long time has quite a low critical temperature $T_c$ extending only to $T_c^{onset}$ = 55 K for the most oxidized sample with $\delta = 0.953$. We find that the $\delta$-dependence of the lattice parameter $c$ significantly changes with time $\tau$. When using the relation between $c$ and $p$, the existence of electron doping of 0.045 per Cu in $Nd_{1.2}Ba_{1.8}Cu_3O_{6+\delta}$ can be inferred. This doping explains well reduced $T_c$ in $Nd_{1.2}Ba_{1.8}Cu_3O_{6+\delta}$. In addition, we observe a local drop of $T_c$ by up to 20 K for the samples with $p \sim 0.095$ ($\delta \sim 0.8$) similar to decreasing $T_c$ known as "anomaly 1/8".

## 1. Introduction

After quenching from oxygen annealing temperatures some properties of the $RBa_2Cu_3O_{6+\delta}$ (R – rare earth element or Y) oxides, in particular, the lattice parameter $c$ and the superconducting transition temperature $T_c$, change for a long time at room temperature (RT) that is known as "room-temperature aging effect" [1–4]. In papers [1–3], these changes in $T_c$ are considered to be connected with the formation of oxygen ordering in the basal $CuO_\delta$ planes at RT, at which copper ions are in the square coordination. According to [1–3], this coordination promotes increasing of the concentration of hole charge carriers in the $CuO_\delta$ planes as well as in $CuO_2$ where superconductivity takes place, Fig. 1, giving rise to growing $T_c$. Relaxation of the lattice parameter $c$ has been examined in detail in our resent work [4] on the $YBa_2Cu_3O_{6+\delta}$ (Y-123) samples with various $\delta$ formed by quenching from different annealing temperatures. It has been revealed that oxygen content dependence of the parameter $c$ transforms from almost linearity to quadratic with time elapsed after quenching $\tau$. This finding has allowed us to suggest a physical model that would connect the quadratic term of the $c(\delta, \tau)$ dependence with the increase of the hole charge carries concentration in Y-123. Thus, the nearest $CuO_2$ planes with charge concentration $q$ (dimensionless unit of charge per one copper ion) interact with each other with a force $F$ proportional to $q^2$, Fig. 1, giving the change in the $c$ lattice parameter by the quadratic law [4]:

$$\Delta c^* \approx \frac{c}{3\varepsilon_0 \sigma^2 \varepsilon_{33}} \cdot q^2 = Ac \cdot q^2, \tag{1}$$

where $\Delta c^*$ is the change of the lattice parameter $c$, occurring due to the electrostatic interaction of $CuO_2$ planes; $\varepsilon_0$ is the vacuum permittivity; $\sigma$ is the elementary area on which the force $F$ acts ($\sigma = a \cdot b$); $\varepsilon_{33}$ is the material elastic constant for the $c$ direction; dimensionless constant $A$ counted from the value $\varepsilon_{33}$ = 186 GPa [5] is equal to 0.345. We note that the model formally does not include any fitting parameters. However, as such, we ought to conventionally consider $\varepsilon_{33}$, since its value varies in different sources, forming an interval. Hence, in the case of application of the model to other members of the $RBa_2Cu_3O_{6+\delta}$ family, coefficient $A$ has to vary around the value of 0.345, reflecting uncertainty in parameter $\varepsilon_{33}$ estimates. According to [4], the change in coefficient $A$ with $\tau$ is negligible.

It is well-known that the superconducting transition temperature of Y-123 is dependent on $q$ [6]. That has given us the possibility to link two experimental functions: $\Delta c^*(\delta, \tau)$ and $T_c(\delta, \tau)$. The experimental checking performed in this manner has showed the efficiency of the developed approach.

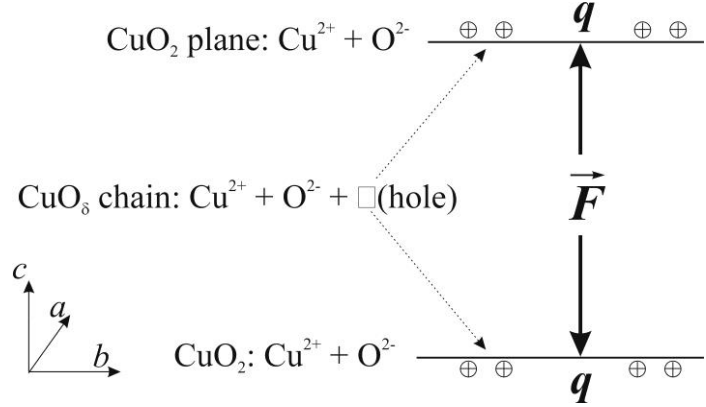

**Fig. 1.** Schematic representation of interacting neighbor $CuO_2$ planes. Dashed arrows show the charge transfer from a $CuO_\delta$ plane (that serves as a charge reservoir) to $CuO_2$.

In the present paper, we use developed model [4] on another member of the same family of superconductors, namely: $NdBa_2Cu_3O_{6+\delta}$. The choice of the neodymium based compound has been dictated by its ability to form a solid solution $Nd_{1+x}Ba_{2-x}Cu_3O_{6+\delta}$ (Nd-123ss) with heterovalent substitution $Nd^{3+}/Ba^{2+}$ in the barium sites [7–13]. It is interesting that with increasing the degree of substitution $x$ in this superconductor, a sharp drop in $T_c$ is observed. Such degradation usually is associated with the appearing of additional oxygen in the structural $CuO_\delta$ plane, causing the breaking of the square coordination of copper [11]. On the other hand, analysis of the total concentration of holes in Nd-123ss [8, 11, 12] has demonstrated that it remains constant and at maximum level till $x \approx 0.3$, with $T_c$ being already much suppressed. This fact casts doubt on the effect of additional oxygen on $T_c$. We consider more possible the drop of $T_c$ in Nd-123ss to be just related to electron doping, which is a characteristic of the systems with heterovalent substitution $Nd^{3+}/Ba^{2+}$ [14].

To check our suggestion we have studied the hole charge carries concentration in the $CuO_2$ planes. This task has been performed with the aid of "aging" method described in [4] – by studying the quadratic term in the experimental dependences $\Delta c^*(\delta)$ obtained at different time $\tau$.

## 2. Experimental details

Solid solution $Nd_{1+x}Ba_{2-x}Cu_3O_{6+\delta}$ (Nd-123ss) with $x = 0.2$ was synthesized from oxides $Nd_2O_3$, $CuO$ and barium carbonate $BaCO_3$, which were mechanically mixed and for a start calcined at 940 °C for the mix to be decarbonized (about 10 h). Then the powder was pressed into pellets and reheated at the temperature 1000°C for 48 h with intermediate regrinding and repelleting. Finally, the material was ground and sintered again at 940°C for 24 h in order to obtain a weak sintered body. As a result, the product with the impurity level of less than 0.5 % was obtained, Fig. 2. A typical Rietveld refinement pattern represented in Fig. 2 was created by using the software package GSAS [15]. Final values of the divergence factors were: $R_{\omega p}$= 8.1%, $R_p$= 5.8%, $R_B$ = 10.6%.

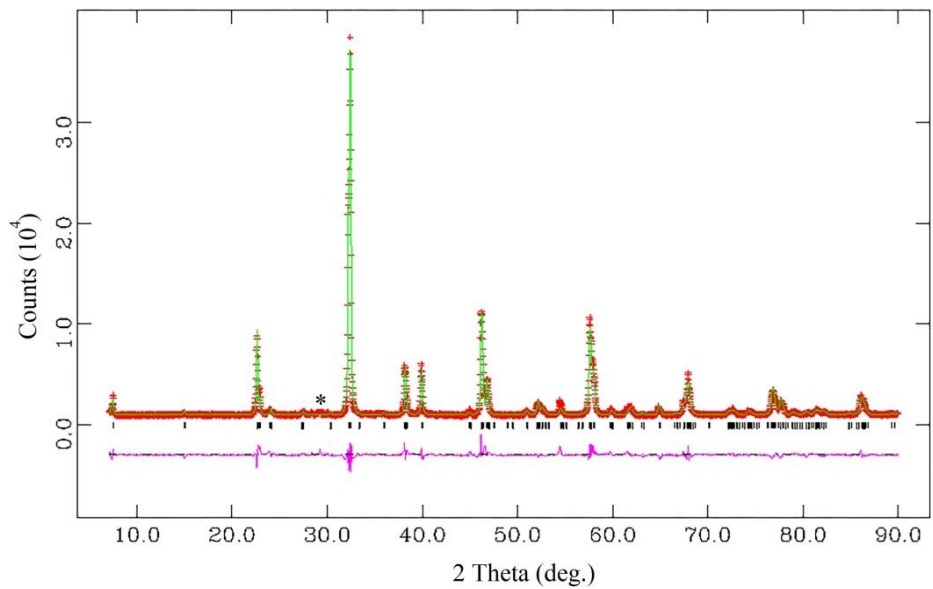

**Fig. 2.** Experimental, calculated and difference diffraction pattern for $Nd_{1.2}Ba_{1.8}Cu_3O_{6+\delta}$ (Nd-123ss); the $CuK_\alpha$ radiation was used in this X-ray diffraction experiment. Position of diffraction lines for Nd-123ss are marked. Under the asterisk there is a signal of the impurity phase $Ba_2Cu_3O_{5+y}$.

To obtain samples with different oxygen content, additional annealing in air at temperatures of the range 470–940°C following by quenching was conducted. More detailed information on this procedure is contained in [4]. The oxygen content was monitored by the change in mass of the sample during additional 1 h -oxidation at 470 °C in air, counting on this treating to give rise to (6+δ) = 6.953 [13].

The samples with different oxygen content were examined by X-ray diffraction (a Shimadzu XRD-7000 diffractometer). Duration of the survey was 25 min; the middle of the survey duration was taken for the calculation of τ. Each sample was examined thrice: at τ = 0.3; 5 and 360 h. Lattice parameters were determined by the least square method, using positions of 14 diffraction lines in the range 2Θ = 20° ÷ 70°. The standard deviation between the calculated and measured positions of the lines was $\overline{\Delta 2\Theta}$ = 0.015°.

The low-temperature magnetic measurements were performed in a vibrating sample magnetometer (VSM) Cryogenic CFS-9T-CVTI, measuring the in-field cooling magnetization at 50 Oe in the temperature range 4.2–150 K with a rate of 1 °C/min.

### 3. Experimental results

#### 3.1. *Study of Nd-123ss by x-ray diffraction analysis*

Fig. 3 shows the oxygen content dependences of lattice parameters *a*, *b*, and *c* obtained through different time intervals τ. The course of the *a* and *b* (δ, τ) curves reflects the existence of the structural transition between tetragonal and orthorhombic phases, occurring in Nd-123ss at δ ≈ 6.6 [13]. Meanwhile, the τ dependence of these parameters is rather weak compared to the same dependence of *c*. It corroborates the result of [4] on the influence of charge in the $CuO_2$ planes namely on the distance between structural layers, that is reflected in equation (1).

We can see that all dependences for all storage times τ have V-shaped singularities at δ ≈ 0.8. Furthermore, besides this singularity, the oxygen content dependence of *c* is nonmonotonic with minimum at δ = 0.78. These are distinctive features of Nd-123ss in contrast, for example, to Y-123 [4] and unsubstituted Nd-123 [16].

#### 3.2. *Study of Nd-123ss by magnetic analysis*

The results of the magnetic study of Nd-123ss samples stored at RT for 360 h are represented in Fig. 4. Fig. 5 gives the values of the superconducting transition temperature (onset) $T_c$ obtained through analysis of the m(*T*) curves. Here it is to be said that the 360 h-exposition of the samples after quenching ought to be considered as an approximation to the limit τ→∞, since the total duration of the aging process, as experienced by many researchers, has never exceeded 15 days.

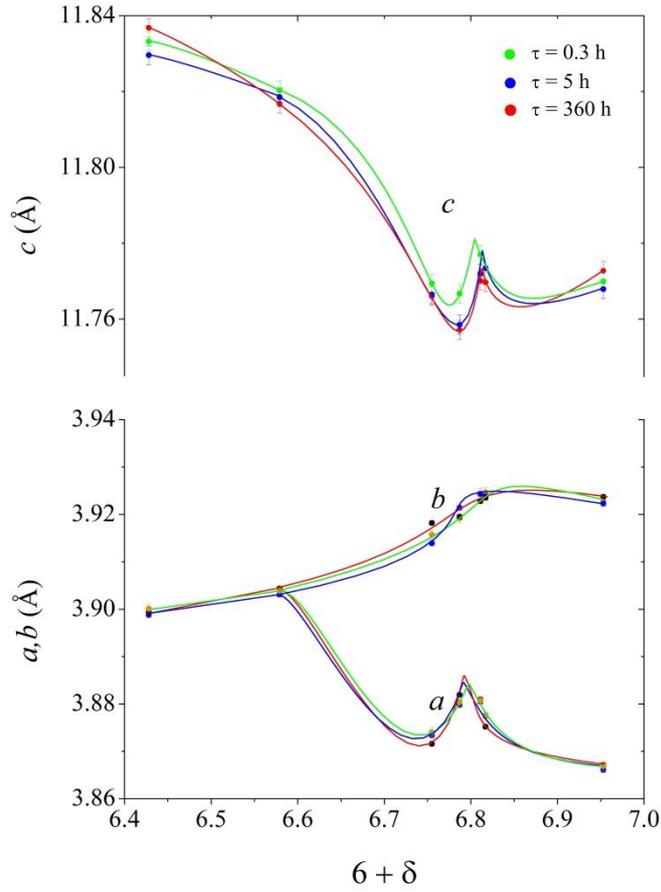

**Fig. 3.** The experimental dependences of lattice parameters *c*, *a*, and *b* on the oxygen content and the time elapsed after quenching of $Nd_{1.2}Ba_{1.8}Cu_3O_{6+\delta}$.

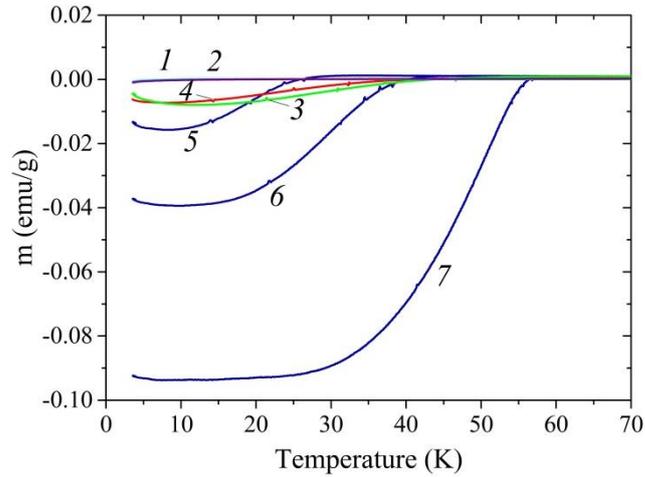

**Fig. 4.** The temperature dependence of the magnetic moment of $Nd_{1.2}Ba_{1.8}Cu_3O_{6+\delta}$. The oxygen content of the samples was: 6.428 (*1*), 6.579 (*2*), 6.755 (*3*), 6.787 (*4*), 6.811 (*5*), 6.817 (*6*), and 6.953 (*7*).

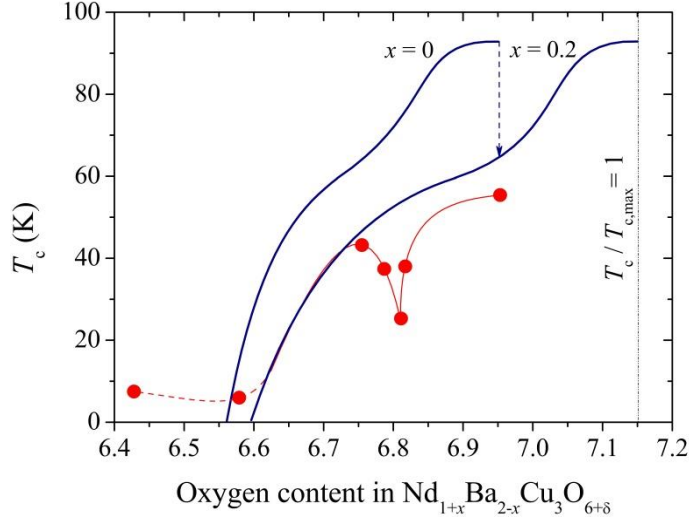

**Fig. 5.** Experimental (red dots with connecting line) and calculated (blue line) data of the superconducting transition temperature as a function of the oxygen content for $Nd_{1.2}Ba_{1.8}Cu_3O_{6+\delta}$. The $T_c(\delta)$ dependence for unsubstituted $NdBa_2Cu_3O_{6+\delta}$ [16] is given for comparison. Dashed arrow shows an expecting change in $T_c$ originated from the substitution $Nd^{3+}/Ba^{2+}$ at $\delta$ = const.

The results of the $T_c$ measurements show that compared to $T_c$ = 94±1 K inherent to Y-123 and unsubstituted Nd-123 [3, 17], the samples of Nd-123ss have quite low critical temperature extending only to $T_c$ = 55 K for the most oxidized sample. However, these are such temperatures that are characteristic for the solid solution $Nd_{1+x}Ba_{2-x}Cu_3O_{6+\delta}$. For example, in the work [9], for similar solid solutions $T_c \approx 80$ (for $x$ = 0.08) and 60 K (for $x$ = 0.16) have been found. More detailed $x$-dependences of $T_c$ are presented in [10–13]. On the basis of them we can evaluate the critical temperature that must be consistent to our oxidized samples with $x$ = 0.2: $T_c \approx 53$ K. This temperature is close to that derived from our magnetic studies.

Fig. 5 shows that while the oxygen content is reducing, a sharp local diminishing of $T_c$ by up to ~20 K occurs near 6+$\delta$ = 6.8, that is very similar to suppressing $T_c$ in $La_{2-x}Ba_xCuO_4$ [18] called "anomaly 1/8". The nature of the latter is known to be the correlated charge and spin ordering in electron-hole system of HTSC at the hole doping of 1/8 per Cu [19–22] that exerts a negative effect on the superconducting properties. On the other hand, coincidence of δ at which the singularities on the $T_c(\delta)$ and $a$, $c$ (δ) curves are manifested can testify about common origin of the observed effects.

## 4. Discussion

Fig. 6 shows difference concentration dependences of the lattice parameter $c$, which visualize the changes in a $c(\delta)$ curve occurring with time after x-ray survey at τ = 0.3 h being made. Represented dependences are obtained from the initial experimental curves (see Fig. 3) by deducing the $c(\delta)$ curve for τ = 0.3 h. The segment of the dependences that corresponds to anomalous reduction of $\Delta c(\delta)$ near δ = 0.8 is depicted by the dashed line. The course of an extracted nonlinear part of difference dependences is shown in Fig. 7. It is namely the dependences that are described by equation (1) (or its modification given below) with accounting for the charge carriers concentration in the $CuO_2$ planes to be a linear rising function of δ [4]. As we can see in Fig. 7, after 360 h the $\Delta c^*(\delta)$ curve looks like the upward parabola, that coincides with the result obtained for Y-123 [4]. This result is predetermined by the gradual acquisition of hole charge carriers in the $CuO_2$ planes with aging. The arising electrostatic forces result in the lattice expansion along the $c$ direction (see Fig. 1).

Meanwhile, at more short aging time, τ = 5 h, corresponding $\Delta c^*(\delta)$ curve looks like the downward parabola that is observed at the first time. According to equation (1), such situation corresponds to a case where the charge carriers concentration in the $CuO_2$ planes decreases with time τ. The only logical explanation for this decreasing is initial (at τ = 0) non-null concentration of doping electron in $CuO_2$. Then, the charge $q$ in equation (1), because of the recombination of electrons and coming holes, will reduce for a while – until the quantities of electrons and holes are equal.

Resently Segawa and Ando [14] demonstrated that in Y-123 with partial substitution of La for Ba the charge balance is provided by both additional oxygen incorporated into the oxide and electron doping of the $CuO_2$ planes. When the transport properties of this solid solution were studied, a smooth transition from $n$- to $p$- type conductivity without the change in the crystallographic structure was observed, that had been anticipated earlier. The transition was adjusted by changing the oxygen concentration in Y-123 and hence by changing the hole concentration in the electron-hole system. Since the solid solution studied in [14] is an analog for Nd-123ss, then the result obtained in the work of Segawa and Ando allows explaining the origin of the initial non-null electron concentration in Nd-123ss as well as the possibility for them to recombine with holes arising in the system. It is worth also to note the paper [23] in which authors demonstrated the electron accumulation in Y-123 films by an electrochemical technique, with *p-n* transition appearing.

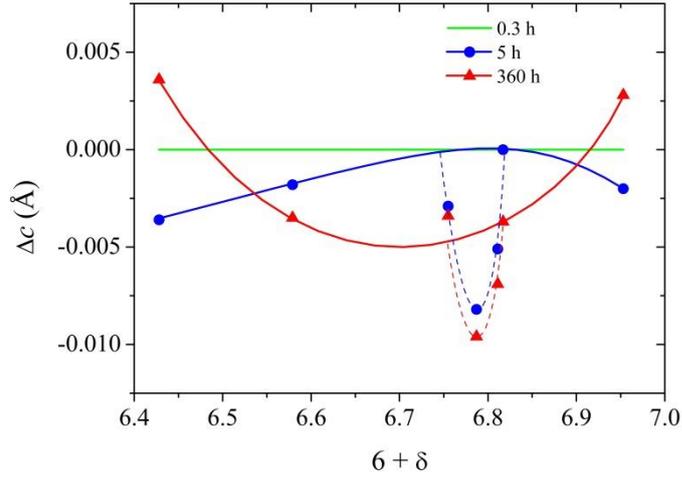

**Fig. 6.** The change of the lattice parameter *c* corresponding to aging of $Nd_{1.2}Ba_{1.8}Cu_3O_{6+\delta}$ from $\tau = 0.3$ h to $\tau = 5$ and 360 h.

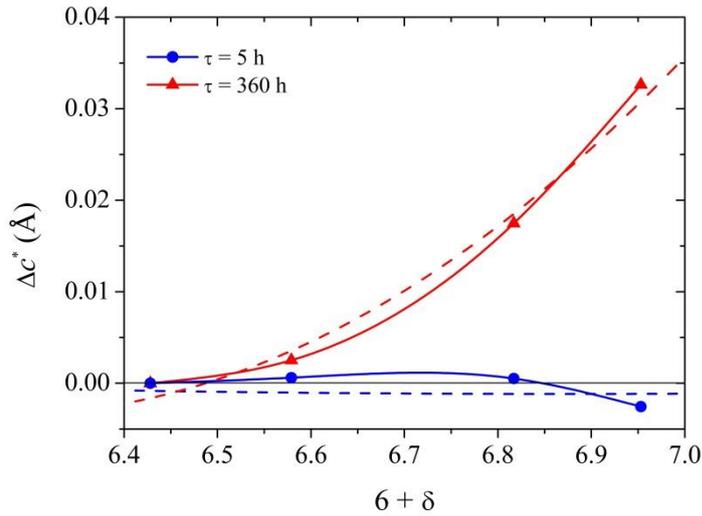

**Fig. 7.** The extracted nonlinear part of the $\Delta c(\delta)$ dependences corresponding to aging time $\tau = 5$ and 360 h. Dots and connecting lines – experiment; dashed lines – calculated data according to equation (3).

To determine the initial electron concentration in Nd-234ss, we further describe the $\Delta c^*(\delta, \tau)$ dependences on the basis of the model presented in [4]. First of all, we have to make the following assumptions: (a) at a given degree of substitution $x$ the degree of electron doping in the $CuO_2$ planes $n$ is a constant; (b) Since the superconducting transition temperatures of Nd-123 and Y-123 in the absence of the $Nd^{3+}/Ba^{2+}$ substitution are very close [16], it would be logical to assume that the relation $p_\infty = 0.187 \cdot \delta$ ($p_\tau$ is the hole concentration corresponding to time $\tau$; $p_{360} = p_\infty$) found for Y-123 [4] is also valid for Nd-123. The $Nd^{3+}/Ba^{2+}$ substitution can be taken into account as follows. Considering the concentration of neodymium ions in barium sites to be $x = 0.2$ and that a part of valent electrons of $Nd^{3+}$ is consumed on doping of two $CuO_2$ planes (till concentration $n$), we find the amount of bonds of oxygen in a $CuO_\delta$ plane with $Nd^{3+}$ to be $x - 2n$. The number of holes that can be formed in $CuO_\delta$ is reduced by the same amount. Then we can write:

$$p_\infty = 0.187 \cdot (\delta - x/2 + n). \qquad (2)$$

In the case of Nd-123ss, parameter $q$ at the time $\tau$ is: $q_\tau = p_\tau - n$. From this we can obtain an expression relating the quadratic term in $\Delta c$ (denoted as $\Delta c^*$) to the formal concentrations of positive and negative charge in the $CuO_2$ planes:

$$(p_\tau - n)^2 - (p_{0.3} - n)^2 \approx \Delta c^* / (A \cdot c); \quad p_\tau = k_\tau \cdot p_\infty, \qquad (3)$$

where $k_\tau$ is the degree of development of the aging effect, which grows from 0 to 1 with time $\tau$. Here $k_\tau$ is the wanted parameter as $n$. Values of $k_\tau$ and $n$ are found by fitting the experimental data in Fig. 4 to equation (4): $n = 0.045$, $k_{0.3} = 0.107$, and $k_5 = 0.187$. Fitting, at which calculated lines $\Delta c^*(\delta)$ (dashed lines in Fig. 7) describe corresponding experimental dots maximum well, is reached at $A = 0.18$. In turn, the concentration of holes coming from the $CuO_\delta$ planes can be expressed now as: $p_\infty = 0.187 \cdot \delta - 0.010$. Note that values $k_\tau$ obtained here, characterizing the aging effect for Nd-123ss, are less than that found earlier for Y-123 ($k_{0.3} = 0.32$ и $k_5 = 0.48$) [4], i.e., relaxation processes occurring after quenching Nd-123ss samples are slower.

In Fig. 5, a dependence $T_c(\delta)$ calculated for Nd-123ss on the basis of empirical relation $T_c(q)$ to compounds R-123 [24] is shown by blue line. Here $q = p - n$ with pa-rameters $p$ and $n$ found above. It is observed that experimental dots $T_c(\delta)$ obtained in our investigation are well described by the calculated dependence, except a small region around $\delta \approx 0.8$ (likely to be connected with "anomaly 1/8"). In the same figure, a dependence $T_c(\delta)$ found for unsubstituted Nd-123 [16] is also presented. By comparing it with the dependence of Nd-123ss we can conclude that sharp drop of $T_c$ resulting from the $Ba^{2+}/Nd^{3+}$ substitution is very likely to be a consequence of shifting the $T_c(\delta)$ dependence to higher $\delta$ and can be explained by the contribution of electron doping in the electron-hole system of the $CuO_2$ planes. Moreover, the presence of the feature on the dependence $T_c(\delta)$ at $\delta \approx 0.8$ allows us to explain data [10–13] on the reducing $T_c$ down to 20–40 K at the substitution level of $x = 0.3$–$0.4$, because further shifting the $T_c(\delta)$ dependence towards high $\delta$ have to result in the emergence of this feature already for the well-oxidized samples.

## 5. Conclusions

Thus, on an example of the representative of the family $RBa_2Cu_3O_{6+\delta}$ (R = Nd) with partial substitution of $Nd^{3+}$ for $Ba^{2+}$ it has been shown that the dependence of the lattice parameter $c$, which is normal to the structural $CuO_2$ planes, on the oxygen content is sensitive to the state of the electron-hole system of these planes. Quadratic term describing the deviation of the $c(\delta)$ dependence from linear function is proportional to the concentration of charge carriers in $CuO_2$ that can be used for a simple experimental evaluation of this parameter in different layered oxides. The concentrations of charge carriers in $Nd_{1+x}Ba_{2-x}Cu_3O_{6+\delta}$ obtained in such evaluations are consistent with the measured superconducting transition temperatures $T_c$.

On the $\delta$-dependence of $T_c$ in the region of a 60 K-plateau, which is characteristic for R-123, a sharp drop of $T_c$ is observed near $\delta = 0.8$. This drop is similar to decreasing $T_c$ known as "anomaly 1/8". The effect requires further experimental investigation.

### Acknowledgments

Authors are grateful to G.A. Dorogina for the help in magnetometry and to S.G. Titova for the fruitful discussion of the results. The investigation was implemented with using equipment of the Center "Ural-M".

Funding: This work was supported by the State Program of IMET UrD RAN [grant no. 0396-2014-0001]